\documentstyle[epsfig]{nature_local}
\newcommand\arcdeg{\mbox{$^\circ$}}%
\newcommand\arcmin{\mbox{$^\prime$}}%
\newcommand\arcsec{\mbox{$^{\prime\prime}$}}%
\newcommand\sun{\odot}%
\newcommand {\gtrsim} {\ {\raise-.5ex\hbox{$\buildrel>\over\sim$}}\ }
\newcommand {\lesssim} {\ {\raise-.5ex\hbox{$\buildrel<\over\sim$}}\ }

\title{A debris disk around an isolated young neutron star}

\author{Zhongxiang~Wang, Deepto~Chakrabarty, \& David~L.~Kaplan
  \institute{Kavli Institute for Astrophysics and Space Research, 
    Massachusetts Institute of Technology, Cambridge, Massachusetts 02139, 
    USA}}

\headertitle{A debris disk around a young neutron star}
\mainauthor{Wang, Chakrabarty, \& Kaplan}

\summary{\noindent Pulsars are rotating, magnetized neutron stars that
are born in supernova explosions following the collapse of the cores
of massive stars.  If some of the explosion ejecta fail to escape, it
may fall back onto the neutron star\cite{col71} or it may possess
sufficient angular momentum to form a disk\cite{md81}.  Such
`fallback' is both a general prediction of current supernova
models\cite{hfw+03} and, if the material pushes the neutron star over
its stability limit, a possible mode of black hole
formation\cite{che89}. Fallback disks could dramatically affect the
early evolution of pulsars\cite{md81,mph01b}, yet there are few
observational constraints on whether significant fallback occurs or
even the existence of such disks.  Here we report the discovery of
mid-infrared emission from a cool disk around an isolated young
X-ray pulsar.  The disk does not power the pulsar's X-ray emission but
is passively illuminated by these X-rays.  The estimated mass of the
disk is of the order of 10 Earth masses, and its lifetime ($\gtrsim 10^6$~yr)
significantly exceeds the spin-down age of the pulsar, supporting a
supernova fallback origin.  The disk resembles protoplanetary disks
seen around ordinary young stars\cite{bsc+90}, suggesting the
possibility of planet formation around young neutron stars.}  \dates{5
August 2005}{21 February 2006}

\begin{document}
\maketitle

The so-called anomalous X-ray pulsars\cite{wt06} (AXPs) are a group of
young ($\lesssim 10^5$~yr) neutron stars with spin periods falling in
a narrow range (5--12~s), no evidence for binary companions, and whose
X-ray luminosities ($\sim 10^{36}$~erg~s$^{-1}$) greatly exceed their
rates of rotational kinetic energy loss ($\sim
10^{33}$~erg~s$^{-1}$). AXPs are generally believed to be
`magnetars'\cite{td96}, which are isolated neutron stars with
exceptionally strong ($\gtrsim 10^{14}$~G) surface magnetic field
strengths and whose magnetic energy ultimately powers their X-ray
emission.  An alternative explanation for AXPs attributes their X-ray
emission to accretion from a residual debris
disk\cite{vtv95,alp01,chn00}, but this model has had difficulties
explaining observations\cite{hkk04}.  The brightest known AXP 
is the 8.7~s pulsar 4U~0142+61, at a distance of 3.9~kpc
(ref.~\pcite{dv06}).  Besides its X-ray emission, the pulsar also has
known optical\cite{hkk00} and near-infrared\cite{hkk04} (near-IR)
counterparts.

As part of a systematic search for fallback disks around young neutron
stars, we observed the field around 4U~0142+61 in the 4.5~$\mu$m and
8.0~$\mu$m bands with the Infrared Array Camera (IRAC) on the Spitzer
Space Telescope to look for the infrared excess predicted by models
for an X-ray heated fallback disk\cite{phn00}.  We found a candidate mid-IR
counterpart at the pulsar's position in both bands (Fig. 1) that has
very unusual IR colours (Fig. 2).  Based on the position coincidence
and colours, we conclude that we have identified the mid-IR
counterpart of 4U~0142+61.

\begin{figure}
\centerline{\psfig{file=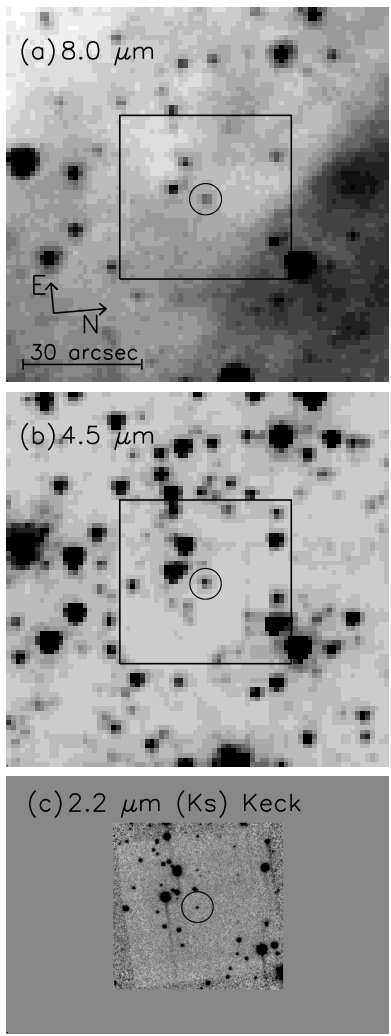,width=2.3in}}
\caption{\small {\bf \,\ \ \ Infrared images of the 4U~0142+61 field.}
  {\bf a, b, } Spitzer/IRAC mid-IR images taken on 17~January 2005 in
  the 8.0~$\mu$m band ({\bf a;} 75~min exposure) and the 4.5~$\mu$m
  band ({\bf b;} 77~min exposure).  An 0.15~arcsec astrometric
  solution was derived using 14 unblended field stars from the 2MASS
  IR point source catalogue.  A candidate pulsar counterpart
  (indicated by a 4~arcsec radius circle) is detected in both IRAC
  images at a position of R.A.~= 01$^{\rm h}$46$^{\rm m}$22.40$^{\rm
  s}$, Decl.~=~$+$61\arcdeg45\arcmin02.9\arcsec (equinox J2000.0),
  with a 1$\sigma$ confidence radius of 0.7~arcsec.  This is
  consistent with the known pulsar position\cite{hkk00}, lying within
  0.3~arcsec.  Based on the source number density in the 4.5 $\mu$m
  image, the chance coincidence probability is only 0.2 percent.  The
  source fluxes were $51.9\pm 5.2$~$\mu$Jy (8.0~$\mu$m) and $36.3\pm
  3.6$~$\mu$Jy (4.5~$\mu$m), where 1~$\mu$Jy=
  $10^{-29}$~erg~cm$^{-2}$~s$^{-1}$~Hz$^{-1}$.  {\bf c,} A deep
  $K_s$-band (2.2~$\mu$m) near-IR comparison image of the central part
  of the IRAC field taken on 1~November 2001 from the Keck~I telescope
  in Hawaii.  (The corresponding subfield is indicated by a box in
  {\bf a} and {\bf b}.)  Besides the pulsar itself ($K_s\approx20$),
  there are no other objects detected within 4.0~arcsec of the source
  position down to a magnitude limit of $K_s\approx 22$.  The scale
  and orientation of {\bf b, c} are the same as shown in {\bf a}.}
\end{figure}

We can reconstruct the observed low-energy spectral energy
distribution of 4U~0142+61 by combining our Spitzer data with existing
IR and optical data\cite{hkk04,isr+04} (Fig. 3).  We may then infer
the intrinsic spectrum by correcting for interstellar reddening.
Although this reddening correction has little effect on the mid-IR data,
it significantly affects the optical data.  For any reasonable choice
of reddening ($2.6<A_V<5.1$; ref.~\pcite{hkk04}), the optical and IR
data clearly arise from two different spectral components.  The
optical (VRI) emission is consistent with a power-law
spectrum\cite{hkk04}, is pulsed at the spin period\cite{km02}, and is
presumably of magnetospheric origin; we will not consider this
component further but will instead confine ourselves to the distinct
IR component.  Hereafter, we adopt the most likely reddening value of
$A_V=3.5$ (ref.~\pcite{dv06}).

We begin by noting that a non-thermal (self-absorbed synchrotron)
origin for the IR component can be ruled out, because associating the
turnover near 4.5~$\mu$m with a synchrotron self-absorption break
requires an emission region with an implausibly weak ($\sim 10^4$~G)
magnetic field strength within a small ($\sim 100$~km) radius of an
active, highly magnetized pulsar.  Similarly, the shape of the
IR spectrum suggests a thermal origin, but the simplest case of a
single-temperature ($\simeq 920$~K) blackbody gives a mediocre fit to
the data.  Moreover, the only plausible blackbody source would be a
large planet or extremely low-mass stellar companion, but the implied
blackbody emission radius ($>5 R_{\odot}$, where $R_\odot$ is the
solar radius) is much too large.

A multi-temperature (700--1200~K) thermal model fits the data better
and thus naturally suggests an extended disk or shell origin for the
IR emission.  We reject a shell geometry as it would produce $\sim
1000$ times the X-ray and optical absorption observed toward the AXP.
As for a disk, it has already been shown that an accretion disk model
(where the pulsar's X-ray flux is entirely powered by disk accretion
onto the neutron star) is ruled out, as it greatly overpredicts the
optical flux\cite{hkk00}.  Instead, we consider a passive disk
illuminated by the X-ray pulsar.  At these low temperatures, the
disk's continuum emission would be entirely from dust.  Indeed, we
note that our observed IR spectrum is quite similar to those of dusty
protostellar disks\cite{bsc+90}, in which the disk is passively
illuminated by the finite-radius central source.  In our case, the
neutron star is effectively an X-ray point source, but assuming there
is enough gas to provide pressure support for the dust, we would
expect any surrounding disk to be slightly flared (with corresponding
disk thickness $h\propto r^{9/7}$ at disk radius $r$;
ref.~\pcite{vrt90}) and thus subject to X-ray irradiation.

\begin{figure}
\centerline{\psfig{file=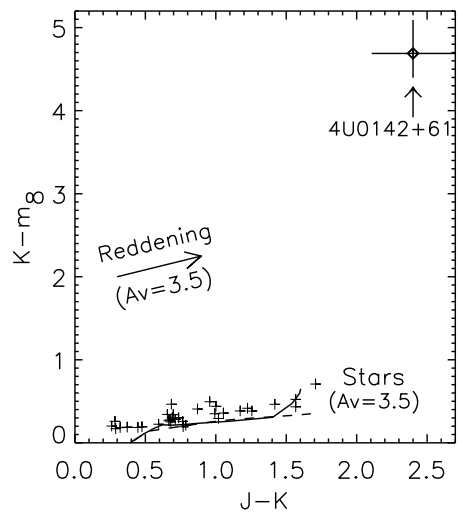,width=3.0in}}
\caption{\small {\bf Infrared colour-colour diagram for 4U~0142$+$61
  (diamond) and 33 field stars (small crosses) from the 2MASS point
  source catalogue.}  We converted the 8 $\mu$m fluxes to magnitudes
  ($m_8$) using a zero point of 63.1~Jy, and used published values of
  the near-IR magnitudes for the pulsar\cite{isr+04}.
  ($K_s=19.9$ is an average value for the pulsar's variable $K_s$-band
  flux\cite{hkk04}.)  The field stars have colours consistent with
  those of reddened ($A_V=3.5$) main-sequence stars (solid line) or
  red giants (dashed lines).  An example $A_V=3.5$ reddening vector is
  also plotted.  The object at the position of 4U~0142+61 has highly
  unusual colours (independent of reddening) and a small chance
  coincidence probability, and we conclude that it is the pulsar counterpart.}
\end{figure}

As additional support for an X-ray--heated disk model for the IR flux,
we note that four of the eight known AXPs have known near-IR counterparts
(including 4U~0142+61), and all of these have similar IR/X-ray flux
ratios of order $10^{-4}$ (ref.~\pcite{dk05}).  In at least one source, there
is even a detected correlation between the near-IR and X-ray
fluxes\cite{tam+04}.  This may indicate that the IR emission
from all AXPs arises in an X-ray--heated debris disk.

\begin{figure}[t]
\centerline{\psfig{file=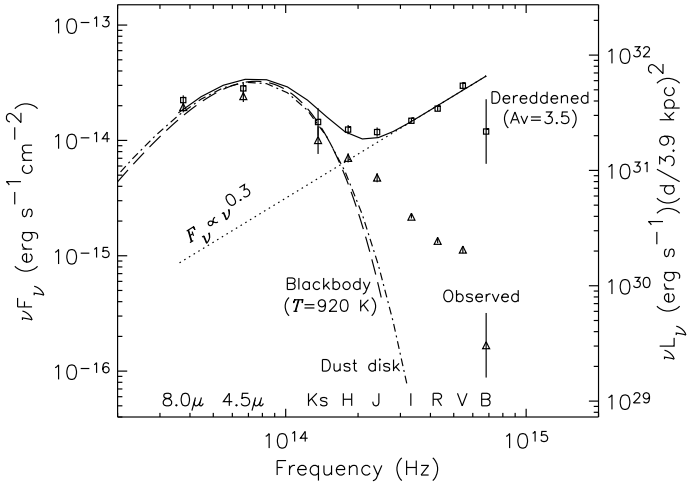,width=3.7in}}
\caption{\small {\bf Optical/infrared spectral energy distribution of
  4U~0142+61.}  The vertical axes are both scaled by frequency
  $\nu$. The left axis shows the $\nu$-scaled flux per unit frequency,
  $\nu F_\nu$; the right axis shows the $\nu$-scaled luminosity per
  unit frequency $\nu L_\nu$ for an assumed distance of $d$=3.9~kpc.
  The $K_s$ point is an average value for the pulsar's variable
  emission in that band\cite{hkk04}.  The triangles indicate the
  observed optical/IR flux while the squares indicate the dereddened
  flux assuming $A_V=3.5$.  The dereddened optical $VRI$ flux is well
  fit by a $F_{\nu}\propto \nu^{0.3}$ power-law component
  (short-dashed line) up to the $B$-band (0.4~$\mu$m) break, and is
  relatively constant in contrast to the near-IR variability
  (0.5~magnitudes).  This power-law is presumably magnetospheric in
  origin, an assertion supported by the optical pulsations that cannot
  be due to X-ray reprocessing\cite{km02}. The power-law--subtracted
  IR spectrum is poorly fit by a simple blackbody model (long-dashed
  curve), likely excluding the presence of large planets or a
  very--low-mass stellar companion.  The best-fit passive debris disk
  model, including irradiation by the central X-ray source, is plotted
  as the dot-dashed curve.  The sum (solid curve) of the disk and
  power-law components fits the optical/IR data (below the break)
  well.  An accretion disk (including irradiation by accretion-powered
  X-rays from the pulsar) can be ruled out because it significantly
  overpredicts the optical flux\cite{hkk00}.}
\end{figure}

We fit the power-law--subtracted IR spectrum from 4U~0142+61 with an
X-ray--heated disk model\cite{vrt90} with 5 parameters: source distance
$d$, disk inclination angle $i$, inner radius $r_{\rm in}$, outer
radius $r_{\rm out}$, and X-ray albedo $\eta_{\rm d}$.  This model
predicts that the disk's effective temperature at radius $r$ is
\begin{equation}
  T(r) \simeq 5,030 \mbox{\rm\ K\ } (1-\eta_{\rm d})^{2/7}
     \left(\frac{d}{\mbox{\rm  3.9 kpc}}\right)^{4/7} 
     \left(\frac{r}{R_\odot}\right)^{-3/7} ,
\end{equation}
where an unabsorbed 0.5--10~keV X-ray flux of $4.8\times 10^{-10}$
erg~cm$^{-2}$~s$^{-1}$ from the pulsar (ref.~\pcite{hkk04}) is assumed
to illuminate the disk.  The total disk flux emitted at a given
frequency is found by integrating the thermal emission over the entire
projected ($\cos i$) disk surface between $r_{\rm in}$ and $r_{\rm
out}$.  For our fits, we fixed $d=3.9$~kpc and $\cos i=0.5$ and
assumed $\eta_{\rm d}>0.9$, as favoured empirically by observations of
X-ray--heated neutron star disks in low-mass X-ray
binaries\cite{dva96}.  We obtained a good fit with $r_{\rm
in}=2.9~R_\odot$, $r_{\rm out}=9.7~R_\odot$, and $\eta_{\rm d}=0.97$.

The inner disk temperature is well constrained to be $T(r_{\rm in})
\simeq 1200$~K by the shape of the near-IR spectrum.  This temperature
is comparable to the sublimation temperature of dust (1000--2000~K,
depending upon grain composition), suggesting that $r_{\rm in}$ may be
set by X-ray destruction of dust.  Disk gas could persist inside
$r_{\rm in}$ without any continuum emission at these temperatures,
cooling instead by line emission.  Alternatively, $r_{\rm in}$ might
also have been set by interactions with the pulsar magnetosphere at
some time in the past, which is consistent with the fact that the
pulsar's light cylinder radius ($r_{\rm LC}=0.61 R_\odot$) lies well
inside $r_{\rm in}$.  The fit value of $r_{\rm in}$ is sensitive to
our choice of $\eta_d$; for $\eta_d =0.5$, the disk is further from
the pulsar with $r_{\rm in}\simeq 19\ R_\odot$, although the fit
quality is poorer.

By contrast, the uncertainty in $r_{\rm out}$ is dominated by our
ignorance of the spectrum at wavelengths longer than 8.0~$\mu$m, since
the location of $r_{\rm out}$ is inferred from the frequency below
which the spectrum turns over to the Rayleigh-Jeans
($F_{\nu}\propto\nu^{2}$) regime.  (Here $F_\nu$ is the flux per unit
frequency at frequency $\nu$.) As this is poorly constrained by
our data, our fitted value for $r_{\rm out}$ should be regarded as a
lower limit.  Observations in the far-IR or millimetre (MM) bands will
be required to determine $r_{\rm out}$ precisely.

Another benefit of far-IR/MM observations is that dust emission in
these bands is typically optically thin, allowing for a direct measure
of the disk mass\cite{bsc+90}. In the absence of far-IR data, we can
at least assume that the flux at 1~mm does not exceed our observed
8.0-$\mu$m flux as the spectrum appears to be falling at
wavelengths longer than 4.5~$\mu$m.  This sets an upper limit on the
total disk mass (both dust and gas) of\cite{bsc+90}
\begin{eqnarray}
 \lefteqn{M_{\rm d} \lesssim 3\times 10^{-3}\ M_{\sun}\, 
    \left(\frac{F_{\rm MM}}{\mbox{\rm 50 $\mu$Jy}}\right) 
    \left(\frac{d}{\mbox{3.9 kpc}}\right)^2} \nonumber \\
 & & \times
     \left(\frac{T(r_{\rm out})}{\mbox{300 K}}\right)^{-1}
     \left(\frac{\kappa_{\rm MM}}{\mbox{0.01 cm$^2$~g$^{-1}$}}\right)^{-1},
\end{eqnarray}
where $M_\odot$ is the solar mass, $F_{\rm MM}$ and $\kappa_{\rm MM}$
are the flux and opacity at 1~mm, and we have assumed a dust-to-gas
ratio of 1\% in the disk.  (Note that our assumptions for
$\kappa_{\rm MM}$ and the dust-to-gas ratio do not account for
the high metallicity and unusual composition possible for
supernova-processed material.)  In protostellar disks, $F_{\rm MM}$ is
generally around two orders of magnitude smaller than the mid-IR
flux\cite{bsc+90}, and this is consistent with an extrapolation of our
best-fit disk model for 4U~0142+61.  This suggests a disk mass of order
$M_{\rm d} \sim 10^{-5} \ M_{\sun}$, comparable to the total mass of
the Earth-mass planets detected around the old millisecond radio pulsar PSR
B1257$+$12 (refs.~\pcite{wf92}, \pcite{wol94}).

In order to consider the lifetime of the disk $\tau_{\rm d}$, we first
need an estimate of the disk's mass loss rate.  For an upper limit, we
can interpret the pulsar's spin period derivative\cite{gk02} ($\dot P=
2\times 10^{-12}$) in terms of magnetic `propeller'
torques\cite{ehn05} to argue that $(-\dot M_{\rm d})\lesssim
10^{-11}\,M_\odot$~yr$^{-1}$. (This presumes that the disk gas
penetrates inside $r_{\rm in}$ and into the pulsar light cylinder to
interact with the magnetosphere at the corotation radius $r_{\rm
co}\simeq 0.01 R_\odot$.)  The resulting disk lifetime $\tau_{\rm d}=
M_{\rm d}/|\dot M_{\rm d}| \gtrsim 10^6$~yr is much longer than the
pulsar's spin-down age $\tau_{\rm p}= P/2\dot P\approx 10^5$~yr,
consistent with a supernova fallback origin.  This also suggests
that the disk did not start out with considerably more mass than it
currently contains.

Our detection of a fallback disk is, to our knowledge, the first
direct evidence for supernova fallback in any context, and it thus
bolsters the notion of fallback in a variety of other circumstances.
For example, direct fallback (without a disk) in the seconds to hours
following core collapse has been proposed as a mechanism for black
hole formation following a supernova explosion\cite{che89}, and may
explain the absence of a detectable pulsar in the nearby supernova
SN~1987A.  However, if a fallback disk forms around a new neutron
star, there are several mechanisms that can limit the disk
lifetime\cite{ph93,mh01}, including accretion, magnetic propeller
expulsion, and ablation by a pulsar wind, with the details depending
upon the spin rate and magnetic moment of the neutron star.  In
addition, giant X-ray flares such as those seen in some
magnetars\cite{hbs+05} may also severely limit the lifetime of
fallback disks.  Our discovery implies that it is possible for a disk
to survive as long as $10^5$~yr, although 4U~0142+61 may be an
exceptional case.

If fallback disks are common around all young neutron stars, then
pulsar spindown may be caused by disk-magnetosphere interactions in
addition to magnetic dipole torques, rendering traditional estimates
of pulsar ages and magnetic moments unrealiable\cite{md81,mph01b}.  On
the other hand, if fallback disks are peculiar to AXPs, this may point
to a distinct formation mechanism for magnetars.  It may also explain
the narrow clustering of slow spin periods for AXPs\cite{alp01,chn00},
despite the need for magnetar activity to account for their X-ray
flux.  However, it is difficult to discriminate among these
possibilities since there are few constraints on the presence of disks
around young pulsars.  Previous searches, motivated by the discovery
of a pulsar planetary system\cite{wf92}, concentrated on middle-aged
and old radio pulsars with no success\cite{lww04}.  Our detection in
4U~0142+61 is probably due to a combination of the X-ray heating of
the disk and the relative youth of the pulsar (compared to $\tau_{\rm
d}$).  We note that energetic particles in a radio pulsar wind may
be as effective as X-rays at heating a disk, so that both X-ray and
radio pulsars are promising targets, and additional debris disks may
yet be detected in a suitably selected sample. However, the strong
non-thermal emission from a young pulsar or supernova remnant may mask
a disk signature in some cases.

By analogy with protoplanetary dust disks around ordinary young
stars\cite{bsc+90}, the presence of a debris disk around a neutron
star naturally raises the possibility of planet formation.  A debris
disk is the presumed origin\cite{ph93,wol94} of the system of
Earth-mass planets already known to exist around the old ($\sim
10^9$~yr) millisecond pulsar PSR~B1257+12, which was detected through
pulsar timing residuals\cite{wf92,wol94}.  However, millisecond
pulsars are generally believed to have been spun up through sustained
disk accretion, and the formation and survival of a debris disk and
planetary system in this context are difficult to explain\cite{mh01}.
Our discovery does not directly address this system, but it does
demonstrate that a debris disk around a neutron star is indeed
possible.  Despite this, it is not clear whether the radiation-rich
environment around 4U~0142+61 is suitable for the formation and
survival of planets\cite{mh01}. On the other hand, since planet
formation probably requires $\sim 10^6$ yr after the disk's
formation\cite{lwb91}, magnetar activity may well have subsided by
that time\cite{td96}.  In any case, searches for planets around young
pulsars are likely to be difficult.  Pulsar timing searches for
planets will have poor sensitivity in AXPs owing to their slow spin
periods, while timing noise in the faster-spinning young radio pulsars
will limit sensitivity to planets in these systems.


\begin{thebibliography}{10}

\bibitem[Colgate<1>]{col71}
Colgate, S.~A. Neutron-star formation, thermonuclear supernovae, and
  heavy-element reimplosion, {\it Astrophys. J.} {\bf 163}, 221--230 (1971).

\bibitem[Michel \& Dessler<2>]{md81}
Michel, F.~C. \& Dessler, A.~J. Pulsar disk systems, {\it Astrophys. J.} {\bf
  251}, 654--664 (1981).

\bibitem[Heger {\it et~al.}<3>]{hfw+03}
Heger, A., Fryer, C.~L., Woosley, S.~E., Langer, N.  \& Hartmann, D.~H. How
  massive single stars end their life, {\it Astrophys. J.} {\bf 591}, 288--300
  (2003).

\bibitem[Chevalier<4>]{che89}
Chevalier, R.~A. Neutron star accretion in a supernova, {\it Astrophys. J.}
  {\bf 346}, 847--859 (1989).

\bibitem[Menou, Perna \& Hernquist<5>]{mph01b}
Menou, K., Perna, R.  \& Hernquist, L. Disk-assisted spin-down of young radio
  pulsars, {\it Astrophys. J.} {\bf 554}, L63--L66 (2001).

\bibitem[Beckwith {\it et~al.}<6>]{bsc+90}
Beckwith, S.~V.~W., Sargent, A.~I., Chini, R.~S.  \& Guesten, R. A survey for
  circumstellar disks around young stellar objects, {\it Astron. J.} {\bf 99},
  924--945 (1990).

\bibitem[Woods \& Thompson<7>]{wt06}
Woods, P. \& Thompson, C. in {\it Compact Stellar X-ray Sources} (eds Lewin,
  W.~H.~G. \& van~der Klis, M.)  (astro--ph/0406133) (Cambridge Univ. Press, in
  the press, 2006).

\bibitem[Thompson \& Duncan<8>]{td96}
Thompson, C. \& Duncan, R. The soft gamma repeaters as very strongly magnetized
  neutron stars. II. Quiescent neutrino, X-ray, and Alfven wave emission, {\it
  Astrophys. J.} {\bf 473}, 322--342 (1996).

\bibitem[{van Paradijs}, {Taam} \& {van den Heuvel}<9>]{vtv95}
{van Paradijs}, J., {Taam}, R.~E.  \& {van den Heuvel}, E.~P.~J. On the nature
  of the `anomalous' 6-s {X}-ray pulsars, {\it Astr. Astrophys.} {\bf 299},
  L41--L44 (1995).

\bibitem[Alpar<10>]{alp01}
Alpar, M.~A. On young neutron stars as propellers and accretors with
  conventional magnetic fields, {\it Astrophys. J.} {\bf 554}, 1245--1254
  (2001).

\bibitem[Chatterjee, Herquist \& Narayan<11>]{chn00}
Chatterjee, P., Herquist, L.  \& Narayan, R. An accretion model for anomalous
  X-ray pulsars, {\it Astrophys. J.} {\bf 534}, 373--379 (2000).

\bibitem[Hulleman, van Kerkwijk \& Kulkarni<12>]{hkk04}
Hulleman, F., van Kerkwijk, M.~H.  \& Kulkarni, S.~R. The anomalous X-ray
  pulsar 4U 0142$+$61: variability in the infrared and a spectral break in the
  optical, {\it Astr. Astrophys.} {\bf 416}, 1037--1045 (2004).

\bibitem[Durant \& van Kerkwijk<13>]{dv06}
Durant, M. \& van Kerkwijk, M.~H. The red clump method for reddening and
  distance determination for the anomalous X-ray pulsars, {\it Astrophys. J.}
  {\bf }, submitted (2006).

\bibitem[Hulleman, van Kerkwijk \& Kulkarni<14>]{hkk00}
Hulleman, F., van Kerkwijk, M.~H.  \& Kulkarni, S.~R. An optical counterpart to
  the anomalous X-ray pulsar 4U0142$+$61, {\it Nature} {\bf 408}, 689--692
  (2000).

\bibitem[Perna, Hernquist \& Narayan<15>]{phn00}
Perna, R., Hernquist, L.  \& Narayan, R. Emission spectra of fallback disks
  around young neutron stars, {\it Astrophys. J.} {\bf 541}, 344--350 (2000).

\bibitem[Israel {\it et~al.}<16>]{isr+04}
Israel, G. {\it et al.} in {\it Young Neutron Stars and Their Environments}
  (eds Camilo, F. \& Gaensler, B.~M.)  247--250 (IAU Symp. 218, Astron. Soc.
  Pacific, San Francisco, 2004).

\bibitem[Kern \& Martin<17>]{km02}
Kern, B. \& Martin, C. Optical pulsations from the anomalous X-ray pulsar
  4U0142$+$61, {\it Nature} {\bf 417}, 527--529 (2002).

\bibitem[Vrtilek {\it et~al.}<18>]{vrt90}
Vrtilek, S.~D. {\it et al.} Observations of Cygnus X-2 with IUE --- ultraviolet
  results from a multiwavelength campaign, {\it Astr. Astrophys.} {\bf 235},
  162--173 (1990).

\bibitem[Durant \& van Kerkwijk<19>]{dk05}
Durant, M. \& van Kerkwijk, M.~H. The broadband spectrum and infrared
  variability of the magnetar AXP 1E 1048.1-5937, {\it Astrophys. J.} {\bf
  627}, 376--382 (2005).

\bibitem[Tam {\it et~al.}<20>]{tam+04}
Tam, C.~R., Kaspi, V.~M., van Kerkwijk, M.~H.  \& Durant, M. Correlated
  infrared and X-ray flux changes following the 2002 June outburst of the
  anomalous X-ray pulsar 1E 2259+586, {\it Astrophys. J.} {\bf 617}, L53--L56
  (2004).

\bibitem[de~Jong, van Paradijs \& Augusteijn<21>]{dva96}
de~Jong, J.~A., van Paradijs, J.  \& Augusteijn, T. Reprocessing of X-rays in
  low-mass X-ray binaries, {\it Astr. Astrophys.} {\bf 314}, 484--490 (1996).

\bibitem[Wolszczan \& Frail<22>]{wf92}
Wolszczan, A. \& Frail, D.~A. A planetary system around the millisecond pulsar
  PSR1257+12, {\it Nature} {\bf 355}, 145--147 (1992).

\bibitem[Wolszczan<23>]{wol94}
Wolszczan, A. Confirmation of Earth-mass planets orbiting the millisecond
  pulsar PSR B1257$+$12, {\it Science} {\bf 264}, 538--542 (1994).

\bibitem[Gavriil \& Kaspi<24>]{gk02}
Gavriil, F.~P. \& Kaspi, V.~M. Long-term Rossi X-Ray Timing Explorer monitoring
  of anomalous X-ray pulsars, {\it Astrophys. J.} {\bf 567}, 1067--1076 (2002).

\bibitem[Ek\c{s}i, Hernquist \& Narayan<25>]{ehn05}
Ek\c{s}i, K.~Y., Hernquist, L.  \& Narayan, R. Where are all the fallback
  disks? Constraints on propeller systems, {\it Astrophys. J.} {\bf 623},
  L41--L44 (2005).

\bibitem[Phinney \& Hansen<26>]{ph93}
Phinney, E.~S. \& Hansen, B.~M.~S. in {\it Planets Around Pulsars} (eds
  Phillips, J.~A., Thorsett, S.~E.  \& Kulkarni, S.~R.)  371--390 (ASP Conf.
  Vol. 36, Astron. Soc. Pacific, San Francisco, 1993).

\bibitem[Miller \& Hamilton<27>]{mh01}
Miller, M.~C. \& Hamilton, D.~P. Implications of the PSR 1257+12 planetary
  system for isolated millisecond pulsars, {\it Astrophys. J.} {\bf 550},
  863--870 (2001).

\bibitem[Hurley {\it et~al.}<28>]{hbs+05}
Hurley, K. {\it et al.} An exceptionally bright flare from SGR 1806$-$20 and
  the origins of short-duration $\gamma$-ray bursts, {\it Nature} {\bf 434},
  1098--1103 (2005).

\bibitem[L{\"o}hmer, Wolszczan \& Wielebinski<29>]{lww04}
L{\"o}hmer, O., Wolszczan, A.  \& Wielebinski, R. A search for cold dust around
  neutron stars, {\it Astr. Astrophys.} {\bf 425}, 763--766 (2004).

\bibitem[Lin, Woosley \& Bodenheimer<30>]{lwb91}
Lin, D.~N.~C., Woosley, S.~E.  \& Bodenheimer, P.~H. Formation of a planet
  orbiting pulsar 1829-10 from the debris of a supernova explosion, {\it
  Nature} {\bf 353}, 827--829 (1991).

\end{thebibliography}

\smallskip
\noindent {\small {\bf Acknowledgements.} We thank M. van Kerkwijk for
sharing the $K_s$-band image of 4U~0142+61.  We also thank A.  Alpar,
L. Bildsten, E. Chiang, M. Durant, E. Dwek, Y. Ek\c{s}i, L. Hernquist,
M. Jura, and R. Narayan for discussions.  This work is based on
observations made with the Spitzer Space Telescope, which is operated
by JPL/Caltech under a NASA contract.  Support for this work was
provided by NASA through a contract issued by JPL/Caltech.  DLK
was supported by a Pappalardo Fellowship.}

\medskip
\noindent {\small {\bf Author information.} The authors declare no
  competing financial interests. Correspondence and requests for
  materials should be addressed to D.C. (deepto@space.mit.edu).}

\end{document}